\documentclass[conference]{IEEEtran}
\IEEEoverridecommandlockouts
\usepackage{cite}
\usepackage{amsmath,amssymb,amsfonts}
\usepackage{algorithmic}
\usepackage{graphicx}
\usepackage{textcomp}
\usepackage{xcolor}
\usepackage{comment}
\usepackage{enumitem}
\usepackage{soul}
\usepackage{url}
\usepackage{hyperref}
\hypersetup{hidelinks}

\usepackage[caption=false,font=footnotesize]{subfig}
\usepackage{tabularx}
\usepackage{booktabs}
\usepackage{adjustbox}
\usepackage{ragged2e}
\usepackage{makecell}
\usepackage{array}
\usepackage{dblfloatfix}

\makeatletter 
\newcommand{\linebreakand}{%
  \end{@IEEEauthorhalign}
  \hfill\mbox{}\par
  \mbox{}\hfill\begin{@IEEEauthorhalign}
}

\def\BibTeX{{\rm B\kern-.05em{\sc i\kern-.025em b}\kern-.08em
    T\kern-.1667em\lower.7ex\hbox{E}\kern-.125emX}}
\begin{document}

\title{Integrating Heterogeneous Digital Twins in Federated Ecosystems}

\author{%
\IEEEauthorblockN{
Christian Vergara-Marcillo\IEEEauthorrefmark{2}\IEEEauthorrefmark{1}\IEEEauthorrefmark{3},
Rami Bahsoon\IEEEauthorrefmark{3}, Nikos Tziritas\IEEEauthorrefmark{4},
\\Wendy Yanez-Pazmino\IEEEauthorrefmark{3}, Panagiotis Oikonomou \IEEEauthorrefmark{4} and 
Georgios Theodoropoulos\IEEEauthorrefmark{1}\IEEEauthorrefmark{2}
}

\thanks{This work was supported by the Research Institute of Trustworthy Autonomous Systems (RITAS), the SUSTech-University of Birmingham Joint PhD Programme, the Guangdong Province Innovative and Entrepreneurial Team Programme (No. 2017ZT07X386), and by the MSc programme in Informatics and Computational Biomedicine of the University of Thessaly. Christian Vergara-Marcillo and Georgios Theodoropoulos are the corresponding authors.}

\vspace{0.15cm}

\IEEEauthorblockA{\IEEEauthorrefmark{1} Research Institute of Trustworthy Autonomous Systems, Southern University of Science and Technology, Shenzhen, China}
\IEEEauthorblockA{\IEEEauthorrefmark{2} Department of Computer Science and Engineering, Southern University of Science and Technology, Shenzhen, China }
\IEEEauthorblockA{\IEEEauthorrefmark{3} University of Birmingham, Birmingham, UK}
\IEEEauthorblockA{\IEEEauthorrefmark{4} University of Thessaly, Lamia, Greece}
}

\maketitle

\begin{abstract}
Digital Twins (DTs) are increasingly used to virtualise physical systems at different scales, enabling monitoring, simulation, and predictions to support decision-making. However, while individual DTs are effective in stand-alone settings, ecosystem-scale deployments require multiple autonomous and distributed DTs to cooperate across system boundaries despite differences in modelling approaches or software technologies, making interoperability and runtime coordination critical challenges. Although \textit{Federated Digital Twin Ecosystems} have emerged as a promising direction, existing research remains at the conceptual stage, offering high-level architectures while leaving the practical integration of heterogeneous DTs underexplored. This paper proposes the \textit{Federation Node Manager}, a modular integration mechanism that connects local DTs to a federated environment through controlled capability exposure, protocol and schema adaptation, and timely state and event exchange for coordinated operations. We present a conceptual design and a prototype implementation, and demonstrate their feasibility in the smart mobility domain for emergency response scenarios. The proposed mechanism serves as an enabling component within a broader service-oriented federated DT ecosystem. 
\end{abstract}


\section{Introduction}

Digital Twins (DTs) are driving innovation across diverse industries by surpassing traditional cyber-physical systems and using high-fidelity models to continuously simulate physical behaviour~\cite{jeong2022digital}. This enables real-time monitoring, optimisation, and predictive analysis for informed decision-making through a seamless feedback loop between the physical and digital domains~\cite{grieves2017digital}. As their adoption expands, stand-alone DTs become insufficient for ecosystem-scale environments that require cooperation across organisational and technical boundaries. This has motivated the notion of \textit{DT Ecosystems}, in which multiple cross-domain twins interact within a shared environment to capture distributed system states and the dynamic interdependencies among connected assets~\cite{ricci2022web}. 

Although these ecosystems facilitate cross-organisational data sharing, system-wide optimisation and planning across spatio-temporal resolutions~\cite{bennett2023towards}, they inherently pose major infrastructural challenges related to heterogeneous DT integration without compromising stakeholder autonomy, data sovereignty and privacy. 

In response, \textit{Federated Digital Twins} have emerged as a paradigm for integrating autonomous DTs into interoperable networks in which multiple participants collaborate toward collective objectives that no stand-alone twin could achieve independently~\cite{10305745}. This paradigm can be understood through the lens of the Federation of Systems (FoS) in System-of-Systems (SoS) engineering, where constituent systems retain \textit{operational and managerial independence} and exhibit high degrees of \textit{autonomy, heterogeneity, and distribution}~\cite{maier1998architecting, sage2001systems}. This allows DTs developed and managed by distinct stakeholders to share selected data and resources without transferring control, thereby preserving local autonomy while enabling richer system-level capabilities. Consequently, this federated approach enables cross-domain analytics, collaborative learning, composite modelling, experimentation, and collective decision-making~\cite{10.1145/3615979.3662152}.

Despite growing interest in federated DTs, the current landscape remains fragmented, often constrained to conceptual architectures, high-level models, or capability-specific solutions~\cite{marah2024re}. As a result, there is limited guidance on how autonomous, heterogeneous DTs can integrate into federated environments while preserving their local sovereignty and stakeholder independence. In particular, interoperability and inter-twin runtime coordination remain major challenges, especially when DTs are conceptualised not only as passive representations (providing only states or telemetry data) but also as active systems capable of local planning and decision-making~\cite{yu2024internet}. The work presented in this paper aspires to address this gap.

In~\cite{10.1145/3615979.3662152} we outlined an architecture for constructing Digital Twins Federations, while in~\cite{10.1007/978-3-031-97632-2_5} we discussed how connectionist theory can underpin the development of such systems. Extending this work, this paper investigates interoperability and runtime coordination in federated environments by proposing the Federation Node Manager as a modular \textit{boundary integration mechanism}. This component enables local DTs to participate in a broader service-oriented federated ecosystem through controlled capability exposure, protocol and schema adaptation, and the timely exchange of states, events and coordination-related artefacts. Grounded in Federation of Systems principles, we present its conceptual design and demonstrate its feasibility in a smart mobility context supporting emergency response use cases. The smart mobility domain is particularly suitable for DT federations, as it involves multiple autonomous systems operating under time-constrained, highly dynamic conditions, while requiring cooperation across heterogeneous DTs that may have been developed separately for distinct operational objectives~\cite{yu2024internet, wang2024augmented}. This is especially relevant in emergency response scenarios where local DTs must exchange evolving, critical context to respond to urgent events while preserving local autonomy for collision avoidance or anomaly detection. 

The contributions of this paper are threefold:

\begin{enumerate}
    \item It frames federated DT ecosystems through core Federation of Systems properties, namely operational and managerial independence, heterogeneity and distribution, deriving autonomy, interoperability and runtime coordination requirements relevant to boundary integration in federated environments.
    
    \item It proposes the Federation Node Manager as a modular integration mechanism that supports federation participation through controlled capability exposure, protocol and schema adaptation, and state- and event-based coordination, enabling heterogeneous DT integration without exposing internal DT logic.
    
    \item It demonstrates the feasibility of the proposed mechanism through a prototype implementation in a smart mobility scenario supporting emergency response use cases.
    
\end{enumerate}

The remainder of this paper is structured as follows. Section~\ref{sec:related_work} reviews related work on DT Ecosystems and ongoing efforts for federated DTs. Section~\ref{fed_autonomy_interoperability_sync} outlines the integration requirements for heterogeneous DTs, which inform the design of the proposed Federation Node Manager detailed in Section~\ref{fdt_federation_node_manager}. Section \ref{sec:prototype_evaluation} validates the mechanism through a smart mobility prototype and emergency response use cases. Finally, Section \ref{sec:conclusions} concludes the paper and outlines future work.

\section{Related Work}
\label{sec:related_work}

As Digital Twins mature, the research landscape has shifted toward ecosystem-oriented and System-of-Systems (SoS) architectures \cite{ricci2022web}. This transition is evident in large-scale initiatives currently developing reference frameworks and standards, particularly for the built environment and smart cities. The UK’s \textit{National Digital Twin Programme (NDTP)} \cite{hetherington2020pathway} proposes the Information Management Framework and the Foundation Data Model to support DT development and interoperability. Similarly, EU initiatives on \textit{Local Digital Twins (LDT)} and Networked LDTs \cite{eu_ldt_toolbox, eu_mim_ldt} promote modular DT ecosystems for cross-organisational integration through shared data spaces, minimum interoperability mechanisms, and semantic technologies. Furthermore, the \textit{Web of DT}s \cite{ricci2022web} proposes an open, dynamic ecosystem of interconnected DTs, supported by knowledge graphs and semantic interoperability. 

In the context of \textit{Federated Digital Twins}, Yu et al. \cite{yu2024internet} propose a hierarchical framework for integrating distributed DTs, emphasising knowledge sharing and cross-domain synchronisation. Similarly, Mahra et al. \cite{marah2024re} detailed a roadmap that includes resource taxonomies and gateway-based integration for horizontal and vertical federations. Other approaches address federation through specific DT capabilities, such as federated learning for cooperative model construction \cite{zhou2023federated}, or multi-layer edge-cloud mobility frameworks that functionally federate vehicles and roadside infrastructure to enhance perception and driving strategies, leveraging reinforcement learning algorithms \cite{wang2024augmented}.

Collectively, the state of the art shows significant progress in federated ecosystem perspectives, semantic interoperability, and domain-specific architectures. However, there remains a critical lack of concrete boundary mechanisms to integrate autonomous DTs for the active exchange of state, events, and decision-related artefacts necessary for collaborative decision-making. Our previous work laid the conceptual foundations for these ecosystems by articulating federations as enablers for autonomous DTs operating in shared virtual environments \cite{10305745}, proposing a high-level architecture for collaborative decision-making \cite{10.1145/3615979.3662152} and conceptualising a connectionist model to govern dynamic inter-twin communication \cite{10.1007/978-3-031-97632-2_5}. Building upon these conceptual bases, this paper bridges the gap to practical implementation by introducing the Federation Node Manager, a boundary integration mechanism that operationalises interoperability, protocol adaptation, and runtime coordination across heterogeneous DTs.

\section{Federated DT Ecosystems Requirements}
\label{fed_autonomy_interoperability_sync}

Federated Digital Twins enable a collaborative network of independent, heterogeneous and distributed DTs~\cite{10305745}. Aligned with the Federation of Systems (FoS) principles~\cite{sage2001systems}, one of their defining characteristics is loose coupling, in which constituent DTs retain \textit{operational and managerial independence}~\cite{maier1998architecting} while contributing to broader system-level objectives. This enables emergent, system-wide capabilities that exceed the benefits of isolated DTs~\cite{marah2024re,10.1007/978-3-031-97632-2_5}. Crucially, federations foster sovereignty and privacy by ensuring that DTs remain locally governed, specialised, and autonomous, and that only selected resources are exposed or exchanged without transferring local control or revealing internal proprietary models and local decision logic.

\subsection{Local DT Autonomy and Sovereignty}
\label{fdt_autonomy}

In federated DT ecosystems, autonomy and sovereignty define the operational boundary between local DTs and the federation. Drawing on autonomy dimensions from foundational federated systems \cite{sheth1990federated}, we interpret local DT autonomy across four dimensions, which determine how capabilities and resources are exposed while preserving local governance. 
\begin{itemize}[noitemsep, topsep=0pt, partopsep=0pt, parsep=0pt]
    \item \textbf{Design Autonomy:} The underlying software, models and decision logic remain internally governed by the stakeholder. Participation relies on selectively exposing resources and capabilities rather than granting access to internal DT components.
    
    \item \textbf{Communication Autonomy:} Local communication and control flows between the physical system and its local twin are separated from federation-level interactions. The federation must support distinct inter-twin interaction patterns without interfering with local processes or disrupting time-critical operations. 
    
    \item \textbf{Execution Autonomy: } DTs retain authority over local operations, allowing them to delay, constrain or reject execution requests in accordance with internal safety constraints or operational priorities.
    
    \item \textbf{Runtime Participation Autonomy:} DTs can dynamically establish, suspend, or withdraw federation interactions based on evolving local context, connectivity conditions (e.g., relevance, trust), or collaboration needs. 
\end{itemize}

\subsection{Interoperability Levels}
\label{fdt_interoperability}

Interoperability enables cross-organisational integration, which is paramount for inter-twin coordination and collaborative decision-making ~\cite{acharya2024interoperability}. In federated ecosystems, it bridges heterogeneous technologies and stakeholder boundaries, enabling mutual information exchange and utilisation~\cite{jeong2022digital, 10305745}. This necessitates a capability that transcends static connectivity to facilitate the active consumption of distributed resources, such as operational states, event streams, simulation endpoints, and decision-related interfaces. Consequently, federations should address six distinct interoperability levels~\cite{acharya2024interoperability} as summarised in Table~\ref{tab:interop_levels}.

\vspace{-1.0em}
\begin{table}[h!]
\centering
\footnotesize
\caption{Interoperability Levels in Federated DT Ecosystems}
\label{tab:interop_levels}
\begin{tabularx}{\linewidth}{p{0.22\linewidth} X}
\toprule
\textbf{Level} & \textbf{Description \& Enabling Technologies} \\
\midrule

\textbf{Technical} & Basic connectivity and reliable data transport (e.g., TCP/IP, MQTT, 5G). \\
\midrule

\textbf{Syntactic} & Readable message structures through common formats (e.g., CSV, JSON, XML, DTDL). \\
\midrule

\textbf{Semantic} & Shared contextual interpretation via common models (e.g., Ontologies, RDF, OWL~\cite{ricci2022web}, FDM~\cite{hetherington2020pathway}). \\
\midrule

\textbf{Dynamic} & Responsive adaptivity under evolving operational conditions (e.g., Microservices, OPC UA, DDS). \\
\midrule

\textbf{Pragmatic} & Alignment of exchanged data with system-specific context and actionable goals. \\
\midrule

\textbf{Organisational} & Alignment of stakeholders, governance models, and policies (e.g., TOGAF, GDPR). \\

\bottomrule
\end{tabularx}

\end{table}
\vspace{-1.0em}

\subsection{Runtime Coordination}
\label{fdt_runtime_coordination}

While interoperability is essential for information exchange, it is insufficient if the data is stale or temporally misaligned. In dynamic federations, enforcing strict global state consistency is often unfeasible, as constituent DTs typically require only localised context (e.g., an autonomous vehicle coordinating with nearby systems rather than the entire city). Instead, these systems rely on runtime coordination to maintain operationally relevant state among active participants. This requires a hybrid approach: periodic time-based updates to maintain \textit{situational awareness}, and event-driven triggers for immediate response to critical conditions. By managing temporal validity and event routing, federated boundary components facilitate dynamic, adaptive cooperation. 

\section{Federated DT Integration: The Federation Node Manager}
\label{fdt_federation_node_manager}

\begin{figure*}[ht!]
  \centering
  \includegraphics[width=0.76\textwidth]{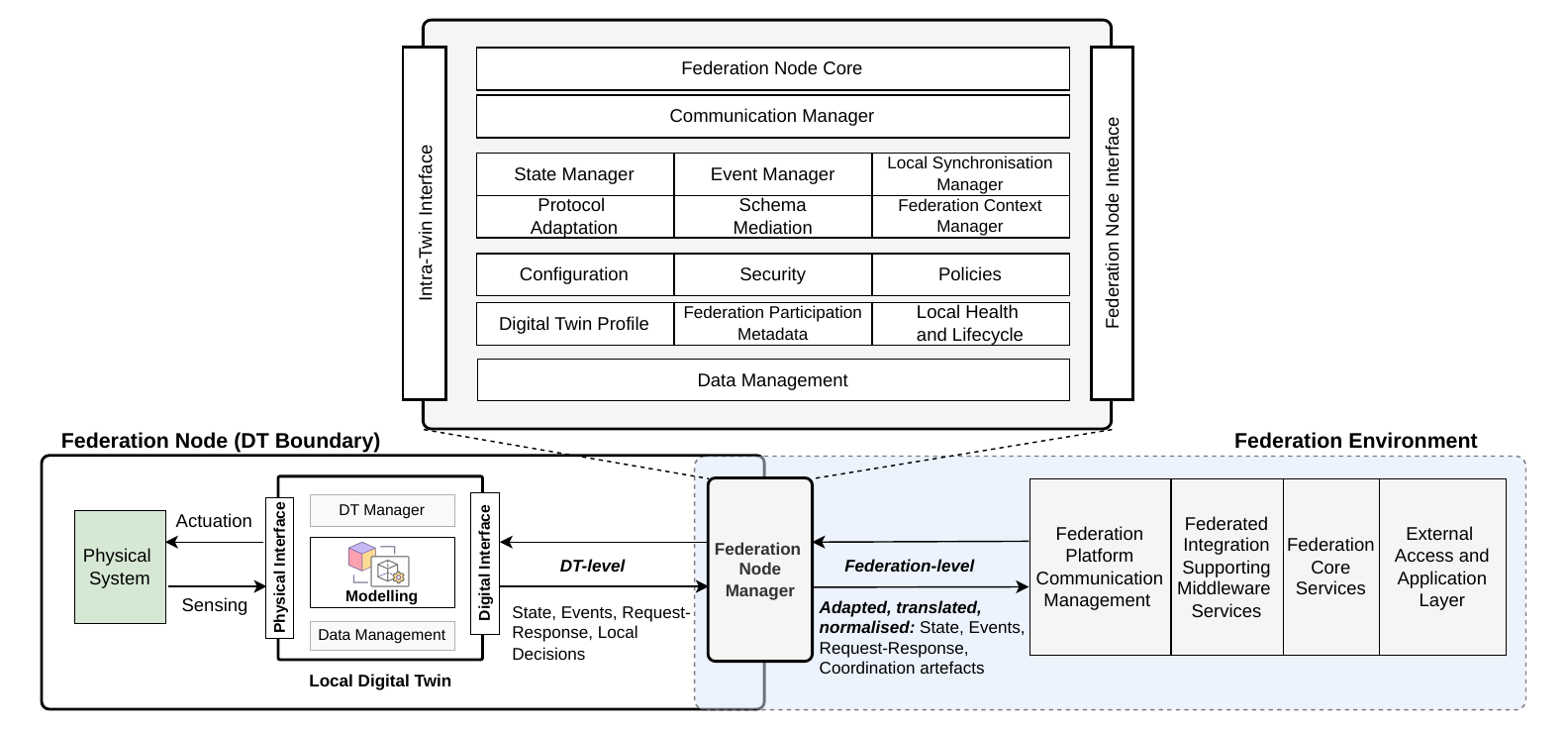}
  \vspace{-0.5em}
    \caption{Architectural overview of the Federation Node Manager. The operational flow (bottom) illustrates this component as the integration boundary between a local DT and a broader federation environment, adapting DT-level artefacts into federation-level interactions. The expanded view (top) details its internal modular architecture, which combines interoperability, runtime coordination support and shared data persistence.}
    \label{fig_fed_node_manager}
    \vspace{-1.5em}
\end{figure*}

Federated DT ecosystems require more than basic inter-twin connectivity to integrate heterogeneous, pre-existing DTs without compromising their autonomy and sovereignty. Since local DT protocols and data schemas inherently isolate these platforms, direct integration of internal DT components is impractical. To address this, we conceptualise each participating DT as a distinct \textit{Node} within a broader collaborative network and introduce the \textbf{Federation Node Manager} as a boundary integration mechanism. Bridging the local DT with a service-oriented federation environment, this unifying abstraction layer adapts and translates local states, events, and other exposed artefacts into a federation-level interaction model. Ultimately, through controlled capability exposure, the Federation Node Manager enables autonomous DTs to collaborate system-wide while preserving local governance, privacy, and operational control (Section~\ref{fdt_autonomy}).

\subsection{Internal Modular Architecture}

Following the requirements derived in Section~\ref{fed_autonomy_interoperability_sync}, the internal architecture of the Federation Node Manager separates orchestration, interoperability, runtime coordination and local governance, relying on shared modules for persistence, metadata, and federation-level handling. As illustrated in Fig.~\ref{fig_fed_node_manager}, a central \textit{Federation Node Core} coordinates five internal concerns to integrate participating DTs without compromising their internal logic:

\begin{itemize}
    \item \textbf{Boundary Interfaces.} The \textit{Intra-Twin Interface} connects to the local DT via operator-configured adapters, while the \textit{Federation Node Interface} connects with the broader federated ecosystem, acting as the local entry and exit point for states, events, and coordination artefacts.

    \item \textbf{Core Control and Orchestration.} The \textit{Federation Node Core} orchestrates internal modules and supervises incoming and outgoing interactions, deciding whether to propagate, cache, defer, or reject federation messages. The \textit{Communication Manager} supports both request-response and publish-subscribe patterns.

    \item \textbf{Interoperability Management.} \textit{Protocol Adaptation} normalises heterogeneous transports (e.g., MQTT, HTTP, or CoAP), while \textit{Schema Mediation} translates local data formats into federation-level schemas or ontologies. Together, they operationalise technical, syntactic, and semantic interoperability at the DT boundary. 

    \item \textbf{Runtime Coordination Support.} The \textit{State Manager} handles periodic state snapshots, while the \textit{Event Manager} supervises discrete events that require prompt federation-wide propagation. The \textit{Local Synchronisation Manager} evaluates temporal validity to ensure the timeliness and relevance of information needed for runtime operations. The \textit{Federation Context Manager} maintains awareness of relevant peers, communication endpoints, and active coordination relationships. 
    
    \item \textbf{Local Governance and Shared Support.} 
    \textit{Security} and \textit{Policies} enforce privacy and sovereignty by restricting access to authenticated systems within the federation. \textit{Configuration} stores operator-defined and local exposure rules to instantiate a Federation Node Manager for a specific application context. The \textit{Digital Twin Profile} and \textit{Federation Participation Metadata} manage identity and discoverability. \textit{Local Health and Lifecycle} monitors DT operational status, while \textit{Data Management} provides a shared persistence layer for caches and federation records.

\end{itemize}

\subsection{Interoperability Support}

The Federation Node Manager operationalises the most immediate interoperability requirements at the DT boundary, as summarised in Table~\ref{tab:fnm_interop_support}. Internal modules enable technical connectivity, normalise syntactic payloads and exchanged data representations, and semantically mediate schemas into a federation-level language. Dynamic interoperability is supported by temporal validity checks and context adaptation under changing conditions. Furthermore, pragmatic and organisational interoperability are only partially supported locally; their full realisation relies on broader governance, operational processes, and trust frameworks at the federation level.

\begin{table}[h!]
\caption{Interoperability support}
\label{tab:fnm_interop_support}
\centering
\begin{tabularx}{\linewidth}{p{0.20\linewidth} X}
\toprule
\textbf{Int. Level} & \textbf{Supported by Federation Node Manager Modules}  \\
\midrule

\textbf{Technical} &
Communication Manager, Intra-Twin Interface, Federation Node Interface, Protocol Adaptation, Federation Context Manager \\
\midrule

\textbf{Syntactic} &
Protocol Adaptation, Schema Mediation \\
\midrule

\textbf{Semantic} &
Schema Mediation, Digital Twin Profile, Federation Participation Metadata \\
\midrule

\textbf{Dynamic} &
State Manager, Event Manager, Local Synchronisation Manager, Federation Context Manager, Local Health \& Lifecycle, Data Management  \\
\midrule

\textbf{Pragmatic, Organisational} &
Partially supported through Security, Policies, Configuration, and Federation Participation Metadata; broader support remains at the federation level. \\
\bottomrule
\end{tabularx}
\vspace{-1.5em}
\end{table}

\addtocounter{figure}{1}

\begin{figure*}[hb!] 
\vspace{-1.5em}
    \centering
    \subfloat[Moderate traffic congestion]{
        \includegraphics[width=0.31\textwidth]{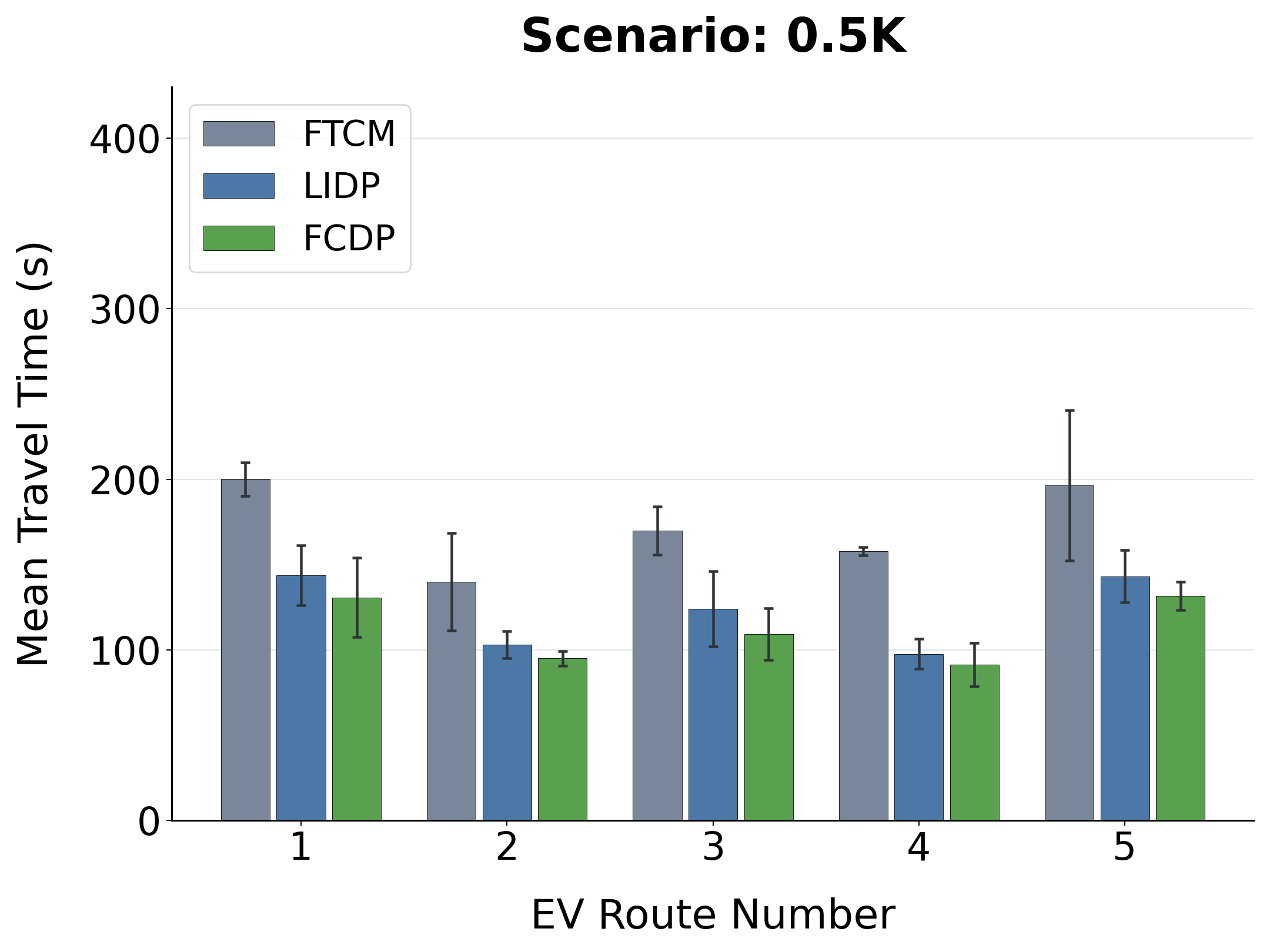} 
        \label{fig:scenario_0_5K}
    }
    \hfill 
    \subfloat[High traffic congestion]{
        \includegraphics[width=0.31\textwidth]{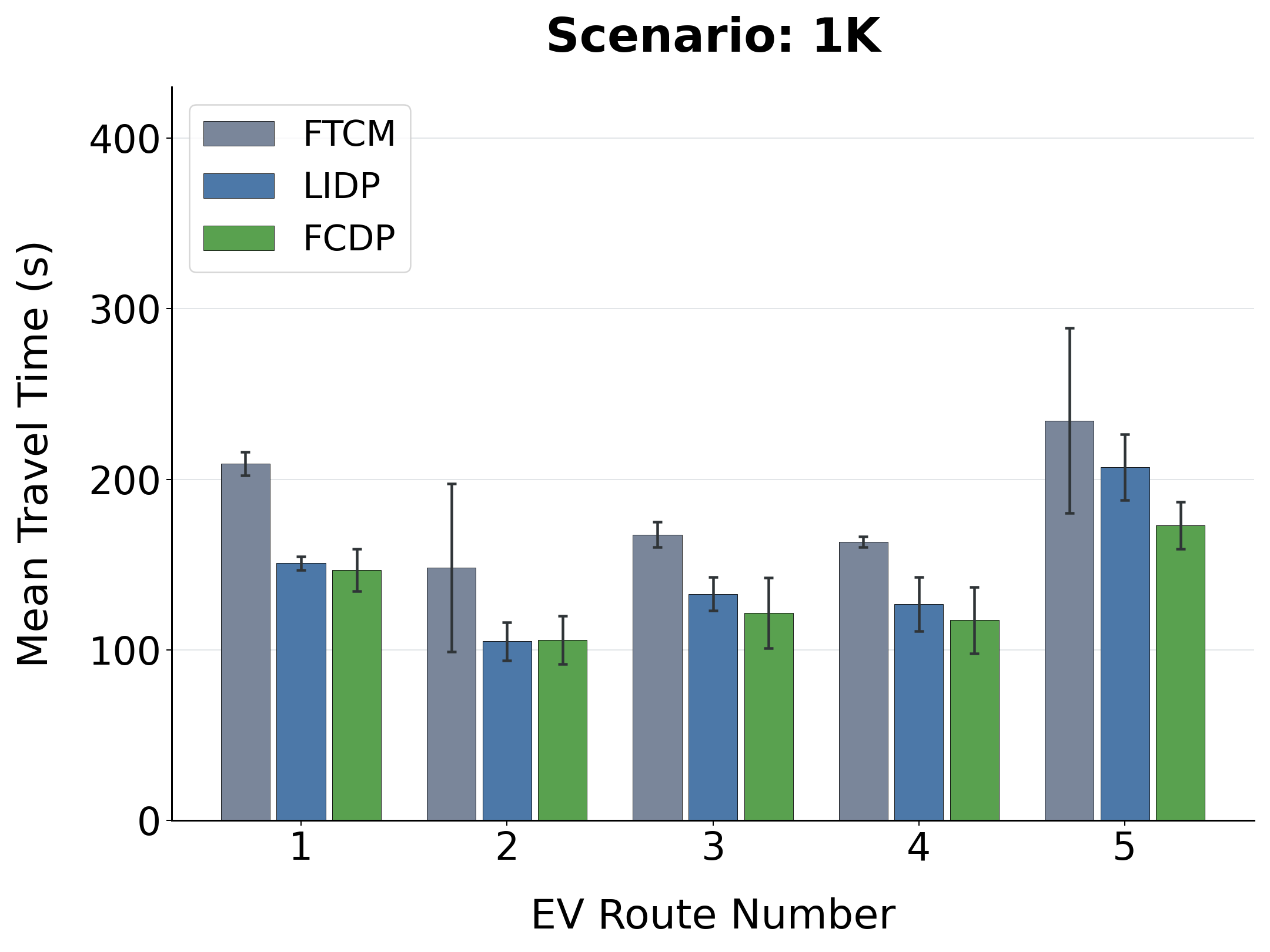} 
        \label{fig:scenario_1K}
    }
    \hfill
    \subfloat[Severe traffic congestion]{
        \includegraphics[width=0.31\textwidth]{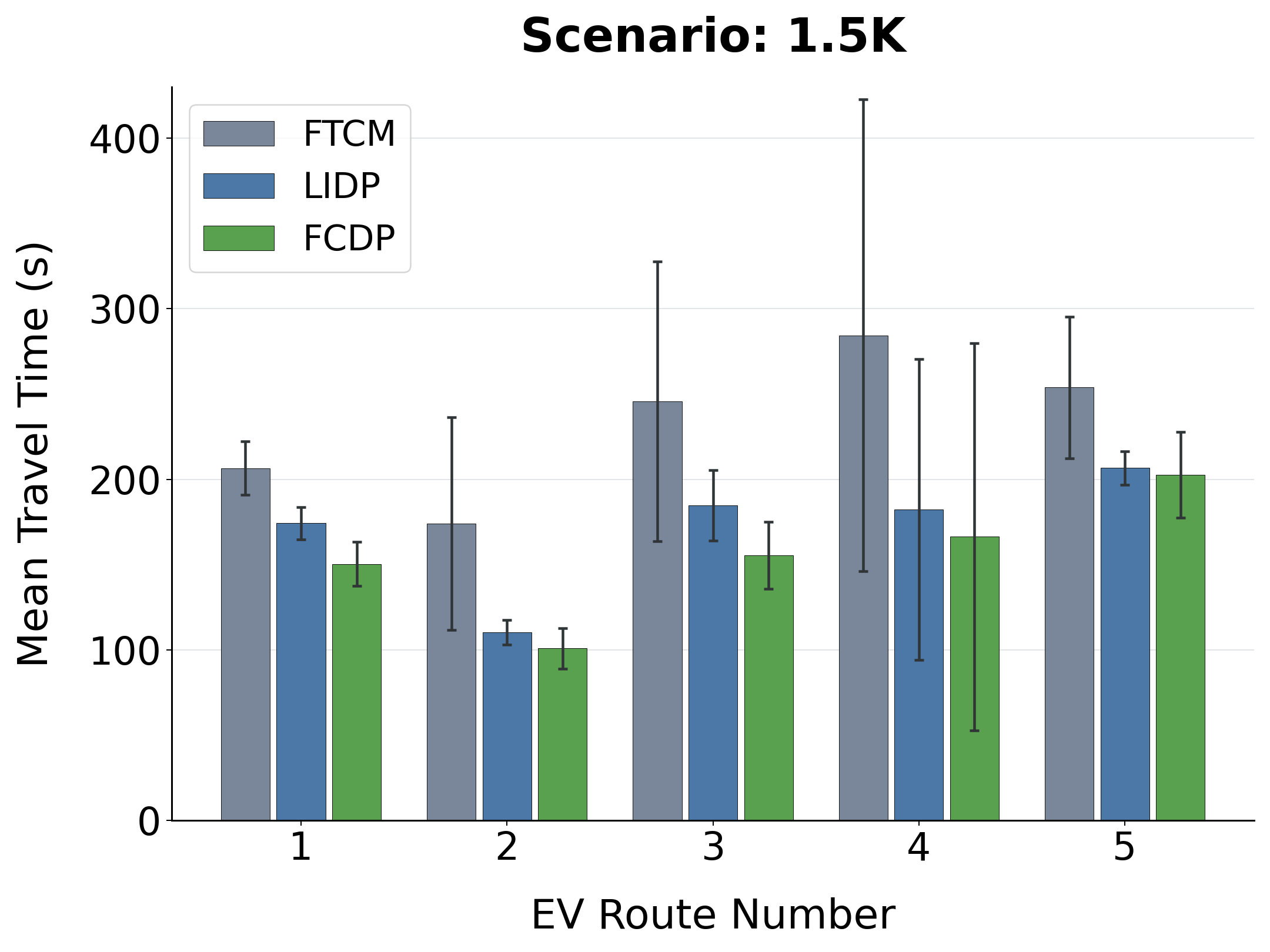} 
        \label{fig:scenario_1_5K}
    }
    \caption{Mean emergency vehicle travel time for five distinct routes. This evaluation compares three operational modes (FTCM, LIDP, FCDP) across Moderate (0.5K), High (1K), and Severe (1.5K) background traffic congestion. Error bars indicate the 95\% Confidence Interval.}
    \label{fig:complete_results}
\end{figure*}

\addtocounter{figure}{-2}
\begin{figure}[ht!]
    \centering
    \subfloat[Federated DT ecosystem and interaction flows]{
        \includegraphics[width=0.95\columnwidth]{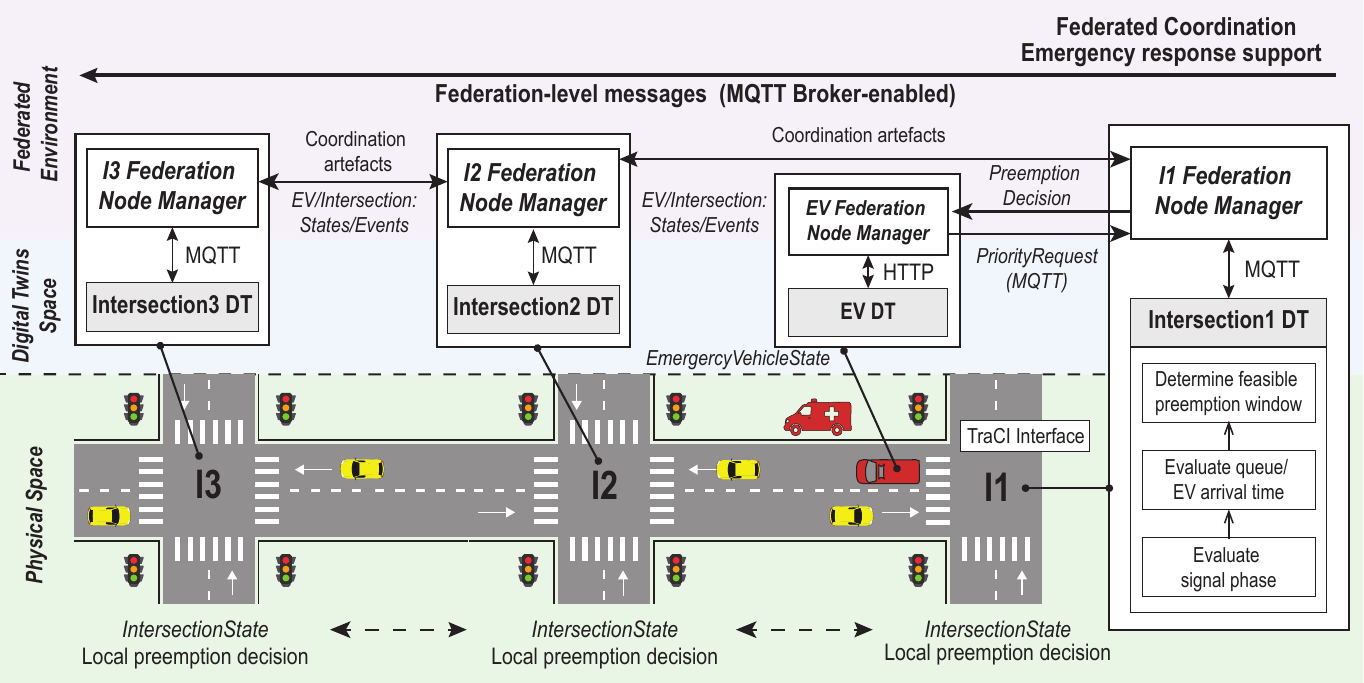}
        \label{fig:sub_traffic}
    }
    \vspace{-0.5em} \\
    
    \subfloat[SUMO traffic simulation snapshot]{
        \includegraphics[width=0.95\columnwidth]{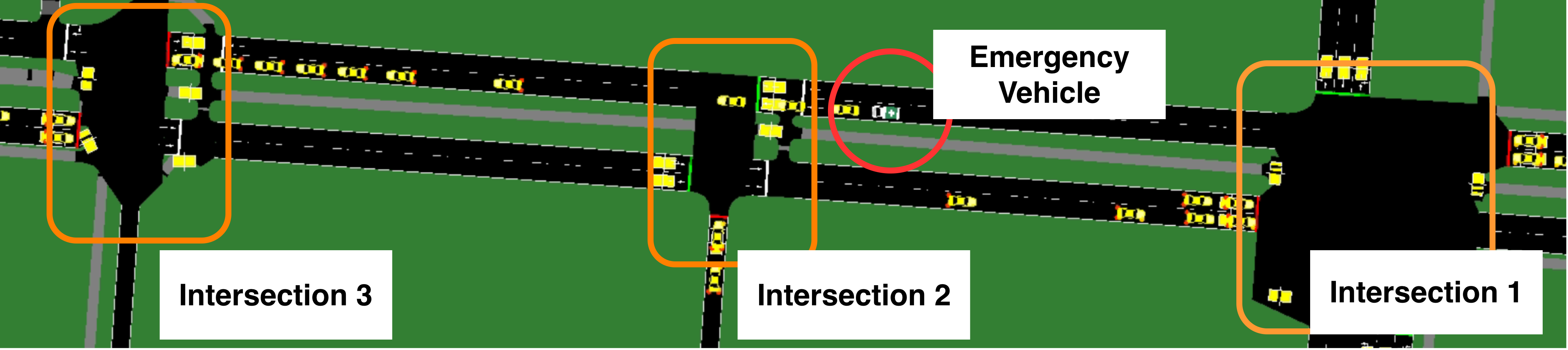}
        \label{fig:sub_emergency}
    }    
    \caption{Federated Digital Twin Ecosystem for emergency response. (a) Architectural overview and interaction flows illustrating the Federation Node Manager enabling inter-twin runtime coordination. (b) SUMO traffic simulation snapshot of the physical environment enforcing the preemption logic.}
    \label{fig:combined_simulation_snapshots}
    \vspace{-1.5em}
\end{figure}
\addtocounter{figure}{1}

\subsection{Runtime Coordination}

Due to heterogeneous update rates and dynamic participation inherent in federated DT ecosystems, enforcing strict global state consistency is often unfeasible. Consequently, the Federation Node Manager addresses coordination as a boundary-level runtime concern, focusing on managing the exchange of timely states and events. To ensure these exchanges remain temporally valid and operationally usable for collaborative decisions, the node prioritises temporal validity over global synchronisation. Supervised by the Federation Node Core, components such as the State and Event Managers, the Local Synchronisation Manager, and the Federation Context Manager utilise timestamp handling, coordination-aware artefacts, validity checks, and local cache management to facilitate context-relevant exchanges among DTs during a specific coordination episode.

\section{Prototype Implementation and Use Case Demonstration}
\vspace{-0.5ex}
\label{sec:prototype_evaluation}

To demonstrate the proposed integration mechanism, we evaluate emergency response scenarios in smart mobility. An emergency vehicle DT (EV-DT) interoperates with multiple smart intersection DTs to enable timely coordination and route preemption. The EV-DT exposes runtime mobility context, including position, lane progress, speed, and estimated arrival time, while each signalised intersection DT executes local preemption logic, alongside federated coordination. This setting is highly suitable for federated integration because isolated local preemption, while improving immediate mobility, often causes downstream spillbacks in constrained corridors; thus, system-wide collaboration and prompt coordination are needed without sacrificing local intersection control.

\subsection{Prototype and Use Case Demonstration}

We implemented a functional prototype to validate the Federation Node Manager as a practical boundary integration mechanism for heterogeneous DTs\footnote{Source code, including message schemas and runtime coordination logic, available at: \url{https://github.com/asia-lab-sustech/fnm-emergency-response.git}}. The prototype utilises the SUMO simulator \cite{8569938} over a constrained urban subnetwork in Madrid, Spain. Traffic dynamics are simulated in SUMO, while intersection preemption logic (adapted from \cite{zhong2022novel}) runs in the DTs and is enforced through SUMO control (Fig.~\ref{fig:sub_emergency}). Each EV-DT and intersection DT is coupled with a Python-based Federation Node Manager that bridges local interfaces (e.g., HTTP) to federation-level MQTT messaging.

The dynamics for state and event exchange, as well as coordination flow, are illustrated in Fig.~\ref{fig:sub_traffic}. The EV-side Federation Node Manager translates continuous \texttt{EmergencyVehicleState} publications into a federation \texttt{PriorityRequest} (\texttt{EVRequest}). The target intersections map incoming artefacts to local interfaces to evaluate preemption feasibility. Furthermore, the Federation Node Manager in the intersections supports inter-twin \textit{runtime coordination} by exchanging \texttt{IntersectionState}, \texttt{CoordinationRequest}, \texttt{SignalWindowProposal} and \texttt{PreemptionDecision} artefacts with downstream peers. Each intersection independently verifies feasibility before actuation, preserving local operational autonomy.

\subsection{Evaluation Methodology}

To quantify the benefits of this federated integration and coordination, we compare three operational modes governing the traffic environment through the intersections: 
\begin{itemize}[noitemsep, topsep=0pt, partopsep=0pt, parsep=0pt]
    \item \textbf{Fixed-Time Control Method (FTCM):} The baseline scenario uses a static, pre-programmed signal timing plan, with no DT intervention or preemption capabilities.
    \item \textbf{Locally Interoperable DT Preemption (LIDP):} A localised integration model, utilising the proposed Federation Node Manager. The EV-DT interacts only with the immediately approaching intersection, with no state sharing or coordination across downstream neighbours. 
    \item \textbf{Federated Coordination DT Preemption (FCDP):} The federated coordination model, utilising the proposed Federation Node Manager. This mode extends EV-to-intersection interoperability by enabling active, downstream inter-intersection coordination. 
\end{itemize}

\subsection{Experimental Results}

Experiments evaluated five random EV routes across three simulated congestion scenarios: moderate (0.5K vehicles), high (1K vehicles), and severe (1.5K vehicles). As detailed in Fig.~\ref{fig:complete_results}, aggregated results across 225 valid traffic simulations indicate a mean EV travel time of 196.74 s for FTCM, 146.13 s for LIDP, and 133.19 s for FCDP. This corresponds to a 25.72\% reduction in travel time when comparing LIDP to FTCM, and a 32.30\% reduction when comparing FCDP with FTCM. Crucially, these results evidence an 8.85\% improvement of FCDP compared with LIDP alone. By congestion profile, FCDP improves over LIDP by 8.75\% at moderate congestion (Fig.~\ref{fig:scenario_0_5K}), 8.00\% at high congestion (Fig.~\ref{fig:scenario_1K}), and 9.64\% in severe congestion (Fig.~\ref{fig:scenario_1_5K}). A paired Wilcoxon signed-rank test on matched runs confirms the statistical significance of these gains ($p<0.001$, Cohen's $d = 0.75$). Supported by 95\% Confidence Intervals (Fig.~\ref{fig:complete_results}), the data confirms that federation-enabled \textit{runtime coordination} delivers concrete end-to-end benefits for time-critical mobility, beyond isolated local interoperability.

\begin{figure}[h!]
\vspace{-1.0em}
  \centering
  \includegraphics[width=0.45\textwidth]{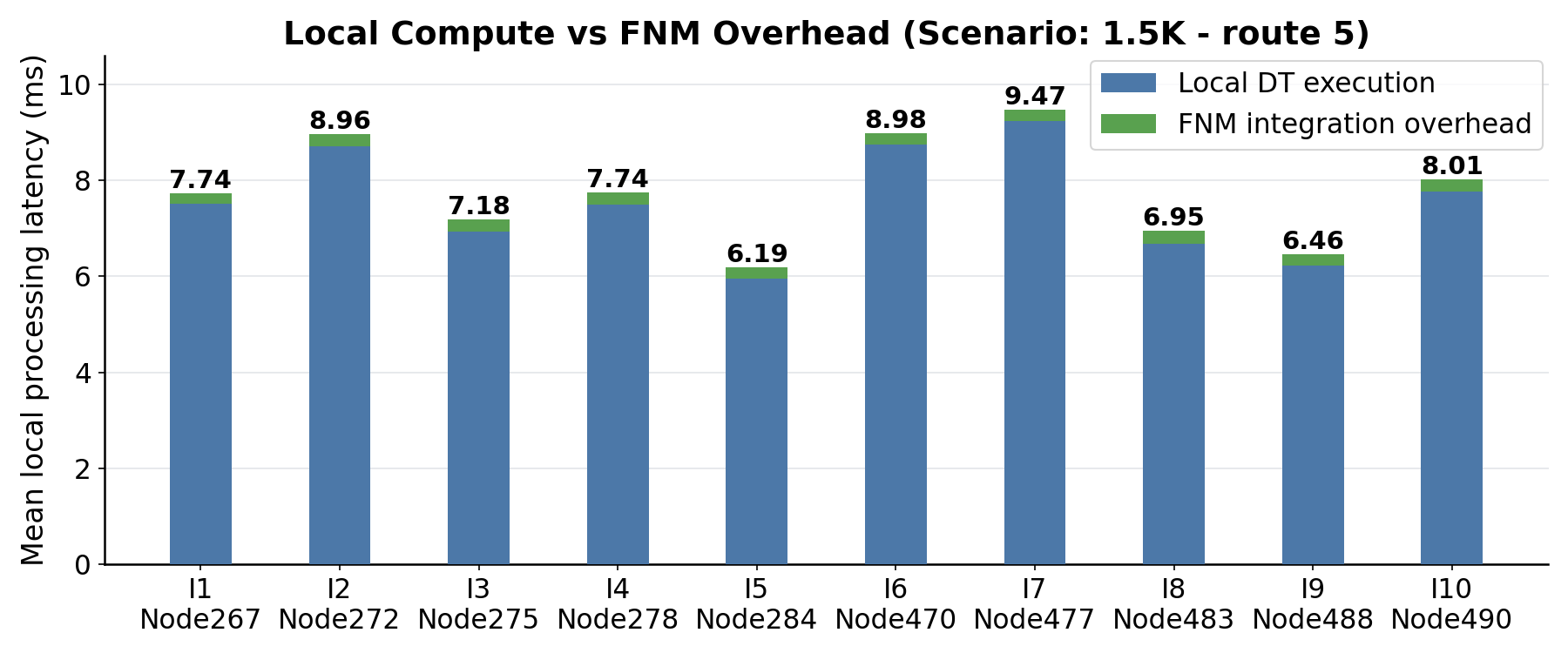}
    \vspace{-1em}
    \caption{Mean local processing latency per node (Scenario 1.5K, route 5).} 
    \label{fig:micro_latency}
    \label{fig:simulation_snapshot}
\vspace{-0.5em}
\end{figure}

Finally, it is critical to ensure that the boundary mechanism does not introduce computational bottlenecks at the edges (at each intersection). Fig.~\ref{fig:micro_latency} reports the mean node-level local processing latency for a representative case (Scenario 1.5K, route 5), separating local DT execution (blue) from the Federation Node Manager integration overhead (green). The results demonstrate that this component adds negligible processing latency: whereas the local intersection DTs remain in the 6.19--9.47 ms range, the added federation overhead is sub-millisecond across nodes. Notably, although network coordination inherently introduces transport and queueing delays (measured between 168.5 ms and 441.5 ms during our simulations, depending on the intersection), this boundary mechanism introduces minimal processing delay. This ensures that federated tasks do not interfere with time-critical local computations, thereby preserving the local operational autonomy (detailed in Section~\ref{fdt_autonomy}) and maintaining the rapid responsiveness required for safe emergency mobility.

\section{Conclusions}
\label{sec:conclusions}
This paper advances the research landscape of Federated Digital Twins, from conceptual frameworks and high-level architectures to practical implementations. Our primary contribution is the Federation Node Manager, a concrete boundary integration mechanism that connects autonomous, heterogeneous and distributed DTs into a federated ecosystem. By enabling controlled capability exposure, protocol and schema adaptation and runtime coordination, this component achieves seamless interoperability while preserving stakeholder sovereignty and encapsulating local DTs logic.

The architecture was validated through a time-constrained smart mobility prototype, demonstrating dynamic runtime coordination between an emergency vehicle DT and multiple smart intersections DTs. The prototype successfully executed timely traffic signal preemptions across distributed boundaries without relying on tightly coupled or centralised monolithic architectures. Serving as a foundational enabler for a broader service-oriented federated DT ecosystem, future work will scale the Federation Node Manager to a complex, intelligent transportation network, evaluate middleware under critical time-constrained conditions, and explore different decision-making topologies across diverse application domains.

\bibliographystyle{ieeetr}
\bibliography{references.bib}

\end{document}